# Introduction of Intellectual Property Courses in STEM Curriculum

Madhur Srivastava
Cornell University, ms2736@cornell.edu

*Abstract* – The paper explains and emphasizes the need to include mandatory Intellectual Property (IP) courses in the undergraduate and graduate curriculum of STEM students. It prescribes the proposed content of IP to be covered at undergraduate level, and graduate depending on the degree pursued by the STEM student. More importantly, this paper advocates for an Integrated PhD/JD degree to produce highly skilled patent attorneys and agents, to overcome the problem of poor quality of patents, and the high number of patent related litigations. A framework is suggested which would enable STEM departments and law schools to successfully offer this integrated degree, and students to successfully pursue and complete this degree program in a realistic manner; without compromising any stake holder's interest.

## INTRODUCTION

STEM enables innovation, invention, discovery, and understanding of technical knowledge which results in products or/and processes in the commercial market leading to the overall welfare of the society. For any innovation or creativity, an individual, a corporation or/and a university has to invest substantial amount of time, money, and other required resources. Therefore, there are Intellectual Property (IP) rights to protect and incentivize innovation, and other forms of creativity. The most important incentives of IP for STEM practitioners are recognition and protection of their work, and economic benefits obtained from the commercialization of their IP.

There are four types of IP rights: trade secrets, copyrights, patents and trademarks, of which STEM practitioners and researchers mostly work in the areas related to copyrights, and patents. However, in most universities, there is no formal curriculum offered or course requirements mandated which teaches IP rights to undergraduate and graduate students in STEM field. Moreover, there is no integrated degree program with enables the teaching of STEM and law at graduate level.

There are two possible occupations for STEM students after their graduation. First, to work on STEM related job, and second, to pursue legal profession in STEM related areas. Usually, a STEM student joins tech-company, tech-consultancy firm, or co-found or be part of a start-up.

Nowadays, a lot of human resources are also utilized in Research and Development (R&D), either in universities through faculty positions, or as research scientists in research labs in the corporations. In all these areas, apart from high technical understanding, jobs require individuals to have awareness of IP rights so that there is no inadvertent IP infringement. Also, it is beneficial to keep in mind IP laws while performing duties for future or prospective innovation protection.

The supernormal growth in the number of patents filed in and granted United States Patent and Trademark Office (USPTO) (see Figure 1; source: [1-2]) suggests that the legal profession needs human resources with sound knowledge of law and technology to evaluate the quality and feasibility of patents. Furthermore, the rise in litigation in patent related cases also demands attorneys and lawyers to possess at least college level STEM education.

Therefore, this paper proposes a curriculum for undergraduate and graduate studies, and an option of an integrated degree in graduate studies incorporating both STEM and law. The aim of the paper is to provide a comprehensive education which mandates some courses in IP law at undergraduate studies; enables IP law courses to graduate students especially looking to work in the tech-companies, start-ups, and R&Ds; and creating a new integrated degree option to STEM members to pursue law.

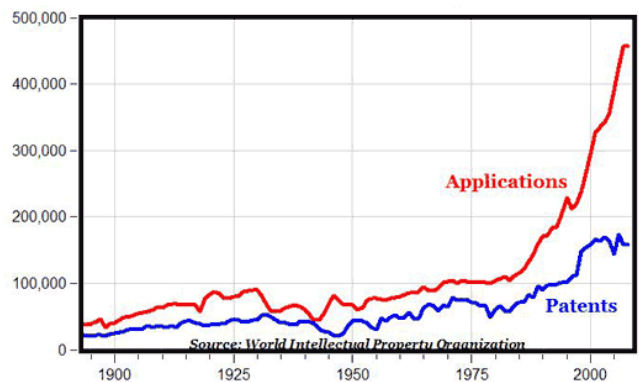

FIGURE 1
U.S. PATENTS APPLICATION AND PATENTS GRANTED: 1883 TO 2008

The paper is organized in the following way. Section 2 illustrates the mandatory IP courses at undergraduate level for STEM students. In section 3, different ways to inculcate IP law among Master's and PhD students, depending on their type of study and degree requirements are provided. Section 3 advocates the need for integrated STEM and law degree. It also proposes a path to obtain such degree in a practical way without compromising either the learning of STEM or law. Finally, section 4 concludes by summarizing the paper.

## IP COURSES AT UNDERGRADUATE LEVEL

The introduction of IP courses at undergraduate level should inform about the following:

1. The illustrative difference between intellectual and physical property.
2. The history, origin and need for IP laws.
3. Types of IP laws, and areas they cover with brief outline of law.
4. The procedure to file an IP, and avenues to protect it.
5. Benefits and drawbacks of IP laws, and economics related to it.
6. Some famous and/or contemporary cases on ongoing IP lawsuits.

The specific number of courses, and credits related to them should be prerogative of the college and its respective departments. Although, the IP laws are much wider and in depth compared to the proposed content suggested to be covered at undergraduate level, the above content in undergraduate studies is comprehensive coverage of all aspects of IP without providing in depth details. The content introduces IP laws in the most general manner. As the career paths chosen are different for different STEM undergraduate students, it would be unnecessary to burden students with specific courses on IP. Further specialization in any specific area of IP laws can be obtained by taking higher level IP courses offered by graduate or law school. Moreover, it would enable students to know how to apply for an IP without getting into too much legality. Lastly, the above curriculum would help a student to choose a career in law involving IP, or a technical career with IP understanding. One of the major advantages of the proposed course curriculum is that the faculty needed to teach the courses could be other than professors from law school.

## IP COURSES AT GRADUATE LEVEL

The graduate students of STEM are pursuing either a Master's degree or a PhD. To simplify, graduate students pursuing STEM degree can be classified into one of the following categories: (1) Master's degree candidate for non-research career, (2) Master's degree candidate for research career, and (3) PhD candidate. Therefore, IP course requirements have to vary for each category.

For category (1) students, the IP course requirements should be same as it is for undergraduate students, except for the fact that the number of courses as well as the number of credits should be reduced to adjust with total course requirements and time commitment. However, all the topics of undergraduate degree IP requirement should be covered. One possible way to achieve this is to offer seminar courses on IP. Also, graduate students belonging to category (1) could also take undergraduate IP courses.

Category (2) STEM graduate students should have a mandatory course requirement of taking 'Intellectual Property' course offered by law school of the home university. In addition, they should have an option to take at least one law school course on specific IP i.e. Patent, Copyright, and Trade Secret, to name a few. If a university or college does not have a law school, then STEM students should be encouraged to attend summer programs on IP offered by different law schools. Further, the choice of seminar courses on IP in addition to above course requirements should be available.

In the case of category (3) STEM graduate students, the IP course requirements should be same as of category (2), except for mandatory two courses on specific IP, instead of one. As PhD students in STEM, research on the areas of innovation, and groundbreaking products and processes, there should be a minor advisor in the PhD committee who is expert in any one of the IP laws. This would serve two purposes: (1) student would conduct his/her research while considering and stating IP laws in his/her field and (2) the work might lead to obtaining IP for the student and research group with the help of the IP expert advisor. Additionally, the IP expert in the committee would benefit from the student's research in analyzing and conducting his/her own research on the effects of IP law in the respective area.

## INTEGRATED PhD/JD DEGREE PROGRAM

There are many scenarios where STEM graduates pursue Juris Doctor (JD) degree to start a career in law mostly related to patents. The rise of patent applications and grants (Figure 1) has resulted in high demand for patent agents and attorneys. Moreover, there is a need for STEM graduates to evaluate the patentability of applications filed in USPTO. To be able to perform either of the above mentioned two jobs, one needs to pass Patent Bar exam.

Recently, the rise of patent litigation, and the number of patents lying worthless, has raised the question regarding the quality of patents which are granted [3]. Therefore, there is need to build a substantial skilled workforce which can understand the innovation, and subsequently, evaluate the innovation's patentability. This workforce needs to be employed at both side i.e. patent filing side as well as patent granting side.

The problem of poor quality patents and the rising number of patent litigation can be analyzed from the educational background and qualification of the people employed in this area. There seems to be two set of educational qualification dominating among the most of the workforce. One pool of employees possesses STEM undergraduate or Master's degree followed by JD. People with this qualification take the job of IP or Patent Attorneys. The other pool of employees possesses just STEM Master's degree or PhD, and takes the job of patent agents. Later, they might pursue JD to become IP or Patent Attorneys. Apart from the above two mentioned qualifications, there are also attorneys who migrated to Patent law. These attorneys may not have had any STEM education.

Even though the educational qualifications of the current workforce related to patents may seem adequate, each pool of employees lack the desired level of technical as well as legal skill to discharge their duties efficiently with minimal faults. The people with STEM undergraduate or Master's degree lack the technical skill to evaluate the quality of innovation because they are not exposed to the amount of skills required to examine the patentability. On the other hand, the patent agents with PhD possess inadequate knowledge of law, especially the awareness of patent lawsuits and their respective judgments. Additionally, the constitutional aspect of patent system is absent among the PhD workforce without JD.

Thus, to overcome the above stated problem of poor quality of patents and high number of litigation associated with patents, this paper advocates the need for integrated PhD/JD degree program which will generate workforce possessing all the necessary skills required to perform their respective jobs.

The PhD degree experience helps the student to obtain expertise in his/her field. Moreover, the comprehension capability of a PhD scholar by reading papers would result in the better understanding of the innovation and its intricacies. This would avoid any duplication of claims w.r.t. other patents while applying for a patent. Furthermore, the technical writing skills acquired when preparing and publishing papers/thesis during PhD would definitely improve the quality in explaining the innovation in a reproducible and organized manner. Both these skills are hard to find among undergraduate and Master's students.

As far as JD degree is concerned, it would provide a holistic legal education of United States and other parts of the world. In addition, IP courses now form an essential part of law degree requirements. It may seem simple to acquire IP law knowledge outside law school, but the cases discussed in IP classes at law schools show that most cases lie in the gray region. In other words, very few infringement cases turn out to be in coherence and accordance with US constitution, common law, and respective IP law, to favor either plaintiff or defendant. One major obstacle is the evolution of technology which may or may not fit into interpretation of any existing laws at that particular time. Hence, reproducible illustration of the innovation while keeping in mind all the related laws becomes important part of patent application. At times, certain claims that are neither supported nor disapproved by any law are filed in patent application. Great legal acumen is required in these types of claims because during litigation, the case can be decided either way. Sometimes, these vacuum are overcome through amendments in law with due recommendations of courts or US congress.

After advocating the need for integrated PhD/JD degree for practicing IP law, more specifically Patent law, following are the three suggestions which would encourage students to pursue PhD/JD in realistic manner.

*I. Single application for admission*

There are many universities offering joint PhD/JD program but none of them (to the best knowledge of the author) have single application for admission in joint degree. A student has to get admission independently in both schools. This can be a major challenge for a student because admission procedure and requirements of law school and graduate school of STEM are different. Therefore, a single application is suggested comprising all the essential admission requirements of both schools. For example, some of the generalized tests can be waived of any school, reducing the preparation time for a student. Further, there can be two Statements of Purpose (SOP): one for STEM graduate school and the other for law school, as an admission requirement. Moreover, as currently applicable to the joint programs in the universities, the option for JD student pursuing PhD or vice versa should be available as long as one is a registered student in any of the two schools.

*II. Special course requirements for integrated PhD/JD degree*

A law school generally requires around 84 credits to be successfully taken to obtain a JD degree (Cornell and Yale law schools). On the other hand, STEM PhD course requirements may vary among universities and departments. It is suggested that joint course requirements do not exceed more than 96 credits (84 law credits and 12 respective STEM credits). This would result in a successful completion of integrated PhD/JD degree in 5-7 years, depending on student's research progress and coursework distribution. The student should be allowed to take any number of credits, except for the first year of study (if required), at any semester of his/her degree candidacy. For example, if a student chooses to complete all the coursework before the start of research, then it would take 3 years to complete it i.e. 16 credits in first 6 semesters. On the other hand, if a student prefers uniform distribution of coursework credits, then he/she would have to take 9.6 (should be rounded-off), 7 and 6.8 (should be rounded-off) credits per semester for the graduation time of 5, 6 and 7 years, respectively. Lastly, there should be minor faculty advisor from law school in the PhD committee. The reasons for this have already been stated in the section 3.

*III. Fully Funded Integrated PhD/JD degree program*

One of the key reasons for not pursuing PhD and JD, sequentially, is the unaffordability of JD tuition. Unlike PhD, STEM students opting for JD after their Master's or undergraduate degree have to arrange high tuition of law school on their own. There are very limit scholarships available in the law school. Therefore, pursuing PhD and JD degree independently is neither time nor monetarily efficient. This problem can be solved through full financial support by university. It can be achieved without any financial or human resource compromise in any school. Firstly, the PhD program is fully funded for most of the students. The tuition of PhD student is either paid by the advisor or the department. There can be a mechanism for tuition sharing between the respective department and law school. It might also be at the cost of reducing the student's stipend marginally. Secondly, there should be limited or fixed number of students enrolled in the integrated program. This would not overburden the law schools with large number of students. Some students from other departments audit law courses and hence, law school can accommodate some extra students. However, the integrated course curriculum should not affect the overall strength of law school i.e. the integrated PhD/JD program students should be additional students, not at the cost of other law students. This solution has advantages discussed in the subsequent paragraphs, for all the students and departments.

Firstly, the law school does not have to forgo any tuition; instead additional students - which it can easily accommodate - would bring more money to law school through tuition sharing.

Secondly, the student's PhD department might have to forgo a part of tuition revenue for only extremely small number of PhD students, which would be negligible, compared to the benefits. The student who applies for the integrated program would be flexible on research to be conducted to obtain a PhD because the ultimate goal of student is to practice law in STEM area. This flexibility would enable faculty members conducting patentable research to include the student in the research group. Moreover, there is also a chance for a research group (previously unaware of patentable subject matter) to apply for patents, which otherwise would not have been thought before the entry of the student pursuing integrated degree. There might be a question that why would a department or a faculty fund a student who does not conduct full time research and uses some his/her time to study law. The answer can be found in the following two advantages for an advisor or a department to fund student: (1) economically, if research group is applying for the patent/s, then the cost of evaluating patentability and preparing draft would be completely forgone due to the student's involvement into it. The consultation fee of attorneys is very high, (2) the student would be most appropriate person to evaluate patentability and write patent draft because of the complete knowledge of innovation. It is very difficult to communicate the innovation to others who have not been part of it, (3) patents are filed for economic benefits through licensing or selling, or opening a start-up company. As the student knows the innovation very closely, finding a prospective licensee becomes easier, and (4) having such student is also time effective in terms of filing patents. Time taken by an attorney, or the patent office of the university is much more (approximately a year), which can be reduced to few months with the presence student actively working on it.

Thirdly, student might receive lesser stipend compared to peers, but in exchange of additional JD degree. This seems better trade-off in comparison to paying the full tuition of law school.

Lastly, the overall duration of program remains between 5 to 7 years. This implies that the advisor or the department do not have to fund student for longer time than what is required for a PhD candidate. Additionally, the JD degree can be accomplished in 5-7 years, contrary to normal 3 years. It means that a student has more time leverage to complete JD. Moreover, law school would not be overburdened to enroll a PhD/JD student in courses that have attained their capacity. The student can take those courses in later semesters giving preference to only JD students.

## CONCLUSION

The paper is divided into two parts: the first part provides the content which needs to be covered to impart the IP knowledge to STEM graduates through different mechanisms, and the second part illustrates the requirement for Integrated PhD/JD degree, mostly to generate skilled patent related workforce. Although, the other forms of IP are not discussed w.r.t. proposed integrated degree program, they can still be pursued under this program, if all stakeholders are convinced with its immediate and future benefits.

The main aim of this paper is to include the legal studies in STEM education. The utility of it should be under educational discourse because as the world is moving towards technical solutions to the world problems, the current educational system should equip the STEM students to deal or avoid with all the legal hindrances related to the advancement of technology for creating welfare states.

## ACKNOWLEDGMENT

The author wants to sincerely thank Parinita Nene (School of Electrical and Computer Engineering, Cornell University) for useful discussions, and providing helpful comments.

## AUTHOR INFORMATION

**Madhur Srivastava,** Department of Biological and Environmental Engineering, Cornell University.